\documentclass[slac_one]{revtex4}
\usepackage{graphicx}
\usepackage{fancyhdr}
\def\DZero{D\O\ }
\def\MET{{\mbox{$E\kern-0.57em\raise0.19ex\hbox{/}_{T}$}}}
\def\met{{\mbox{$E\kern-0.57em\raise0.19ex\hbox{/}_{T}$}}}

\def\DZero{D\O\ }

\def\ifb{~fb$^{-1}$}
\def\gevc{~GeV/c}
\def\gevcc{~GeV/c$^2$}
\def\pp{$p\bar{p}$}

\def\tt{$t\bar{t}$}

\def\lmet{$WH\rightarrow \ell\kern-0.45em\raise0.19ex\hbox{/} \nu b\bar{b}$}

\def\hww{$H\rightarrow W^+ W^-~$}
\def\hwwll{$H\rightarrow W^+ W^- \rightarrow \ell^\pm \ell^\mp + \etmiss$}
\def\hwwlnujj{$H\rightarrow W^+ W^- \rightarrow \ell \nu jj$}
\def\vbf{$q\bar{q} \rightarrow q\bar{q} H \rightarrow qqW W~$}

\def\tevE{$\sqrt{s}=1.96$~TeV}

\newcommand{\Eslash}{\mbox{$\rm E \kern-0.6em\slash$                }}
\newcommand{\etmiss}{\mbox{$\rm \Eslash_{T}\!$                        }}

\newcommand{\mh}{\mbox{$m_H$}}

\newcommand{\wjets}{\mbox{$W+jets~$}}

\begin{document}

\title{High Mass Higgs Boson Searches at the Tevatron} 

%

\author{Bj\"orn Penning for the \DZero and CDF Collaboration  \\
 Fermi National Accelerator Laboratory\\}
\affiliation{FNAL, Batavia, IL 60510, USA}

\begin{abstract}
We present results from CDF and D\O\ on direct searches for high mass standard model (SM) Higgs boson ($H$) in \pp~collisions at the Fermilab
Tevatron at $\sqrt{s}=1.96$~TeV. Compared to previous Higgs boson Tevatron combinations, more data and new channels (\hwwlnujj,
$H \rightarrow WW \rightarrow \ell\tau+X$ and trilepton final states)  have been added.  Most previously used channels have been
reanalyzed to gain sensitivity. Analyzing 5.9\ifb~of data at CDF, and 5.4-6.7\ifb\ at D\O, the combination 
exclude with 95\% C.L. a standard model Higgs boson in the mass range of $m_{H}=$158-175\gevcc.
\end{abstract}

\maketitle

\thispagestyle{fancy}


\section{INTRODUCTION} 

The search for the Higgs boson is one of the main goals of High Energy Physics. In this proceedings we report an update of the searches performed by the \DZero and CDF collaboration for a Higgs boson with a large mass ($>\sim~140$\gevcc). The data has been collected at Tevatron collider at $\sqrt{s}=1.96$ TeV. 
The main search channel in this mass range is the decay mode \hwwll with $\ell=e,\mu, \tau$.  A new analysis utilizes the final state $H \rightarrow W^\pm W^\mp \rightarrow \ell \nu jj $.  While these channels utilize mainly the gluon- and vectorboson-fusion production modes the associated production leads to $W/Z+H \rightarrow W/Z+WW \rightarrow \ell^\pm \ell^\mp(\ell) + X$ with either a same-sign lepton pair or three leptons in the event.

\section{EXPERIMENTAL ENVIRONMENT}

The Tevatron is a \pp~collider with an center of mass energy of \tevE~hosted at the Fermi National Accelerator Laboratory (Fermilab).
Currently more than 9\ifb~of data have been delivered to the CDF and \DZero detectors. Results presented in this proceedings are based on data samples with integrated luminosities from 5.4 to  6.7\ifb. Both, the CDF and \DZero experiment, are multi-purpose detectors. The coordinate system used is the azimuthal angle $\phi$ and the pseudo-rapidity
$\eta=-ln\left(\tan{\frac{\theta}{2}}\right)$. $\Theta$ is the polar angle defined relative to the proton beam axis.
The \DZero detector has a central tracking system which consists of a silicon micro-strip tracker and a central fiber tracker, both located within a 2 T axial magentic field. The tracking system is surrounded by a liquid-argon/uranium calorimenter. Finally muons are identified by detectors comprising of layers of tracking detectors and scintillators in a 1.8 T toroid magnetic field. 
CDF tracking detectors consist of a silicon micro-strip detector array surrounded by a cylindrical drift chamber in a 1.4 T axial magnetic field. Outside of the tracking chambers, the energies of electrons and jets are measured with segmented sampling calorimeters; the outermost detectors are layers of steel instrumented with planar drift chambers and scintillators used for muon identification.

\section{HIGH MASS HIGGS BOSON SEARCHES AT THE TEVATRON}

\subsection{Higgs Boson Production and Decay at the Tevatron}
There are three main production mechanisms for Higgs bosons at the Tevatron collider.
The signal is produced in association with vector bosons ($q\bar{q} \to W/ZH$), gluon-fusion ($gg \to H$) or vector boson fusion ($q\bar{q} \to q'\bar{q}'H$). At low masses $\sim < 140$ the Higgs boson will dominantly decay via $b\bar{b}$ pairs. Due to the overwhelming multijet (MJ) background the large gluonfusion production mode cannot be accessed. Here the associated production $VW$ with $V=W,~Z$ is most important since the accompanying gauge boson can be used to reduce the backgrounds. For higher Higgs boson masses the dominant decay takes place via two $W$ bosons, allowing to take advantage of the larger gluon-fusion production mode and providing clean finals states such as the \hwwll signature. Therefore the high mass Higgs boson searches at the Tevatron offer currently the world's best sensitivity for a standard model Higgs boson. Figure~\ref{fig:prod} shows SM Higgs boson production mode at the Tevatron and the decay for various Higgs boson masses.

\begin{figure}[h!]
\centering
  \begin{minipage}{0.49\textwidth}
    \centering
    \includegraphics[scale=0.32, angle=-90]{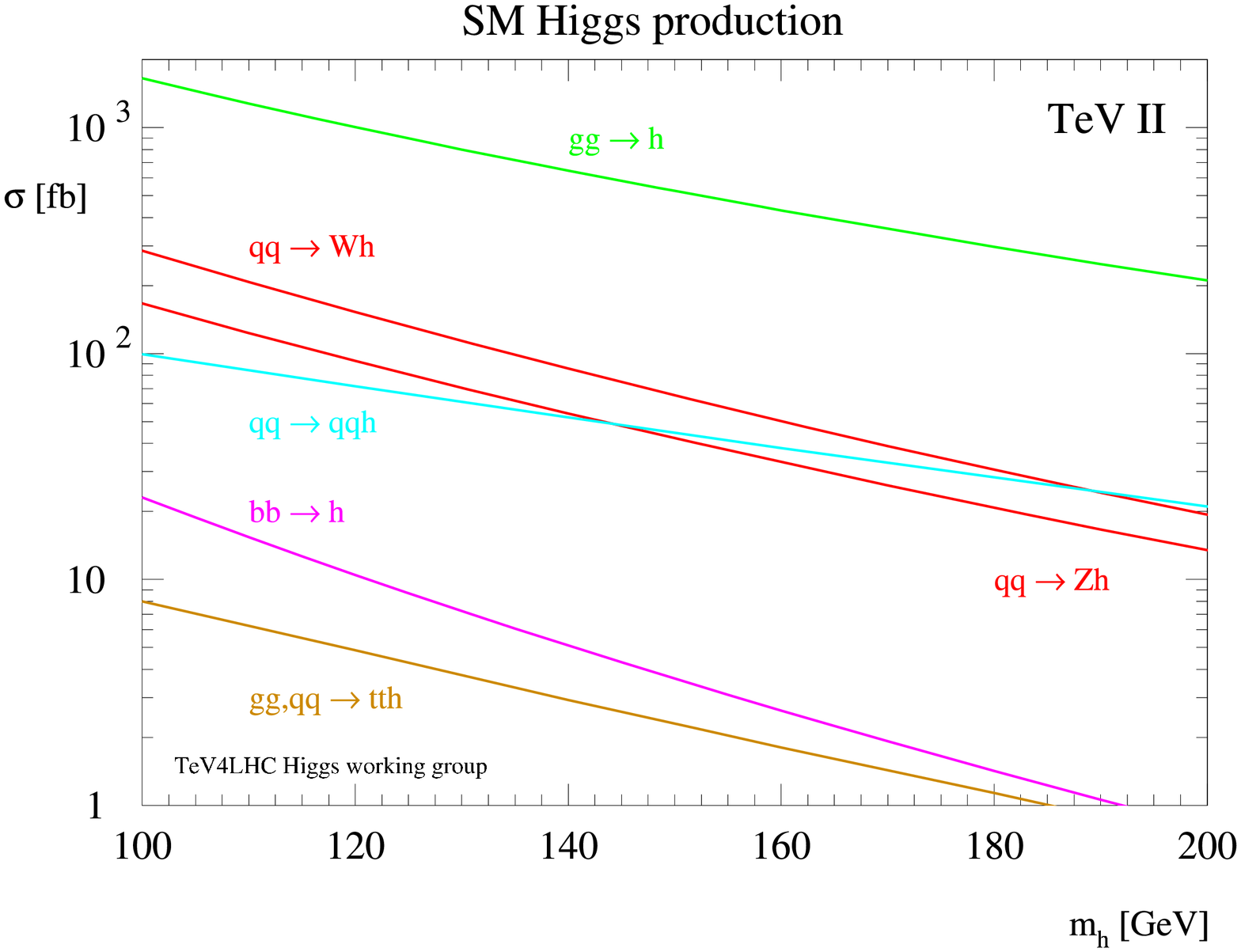}
  \end{minipage}
  \hfill
  \begin{minipage}{0.49\textwidth}
    \centering
    \includegraphics[scale=0.4]{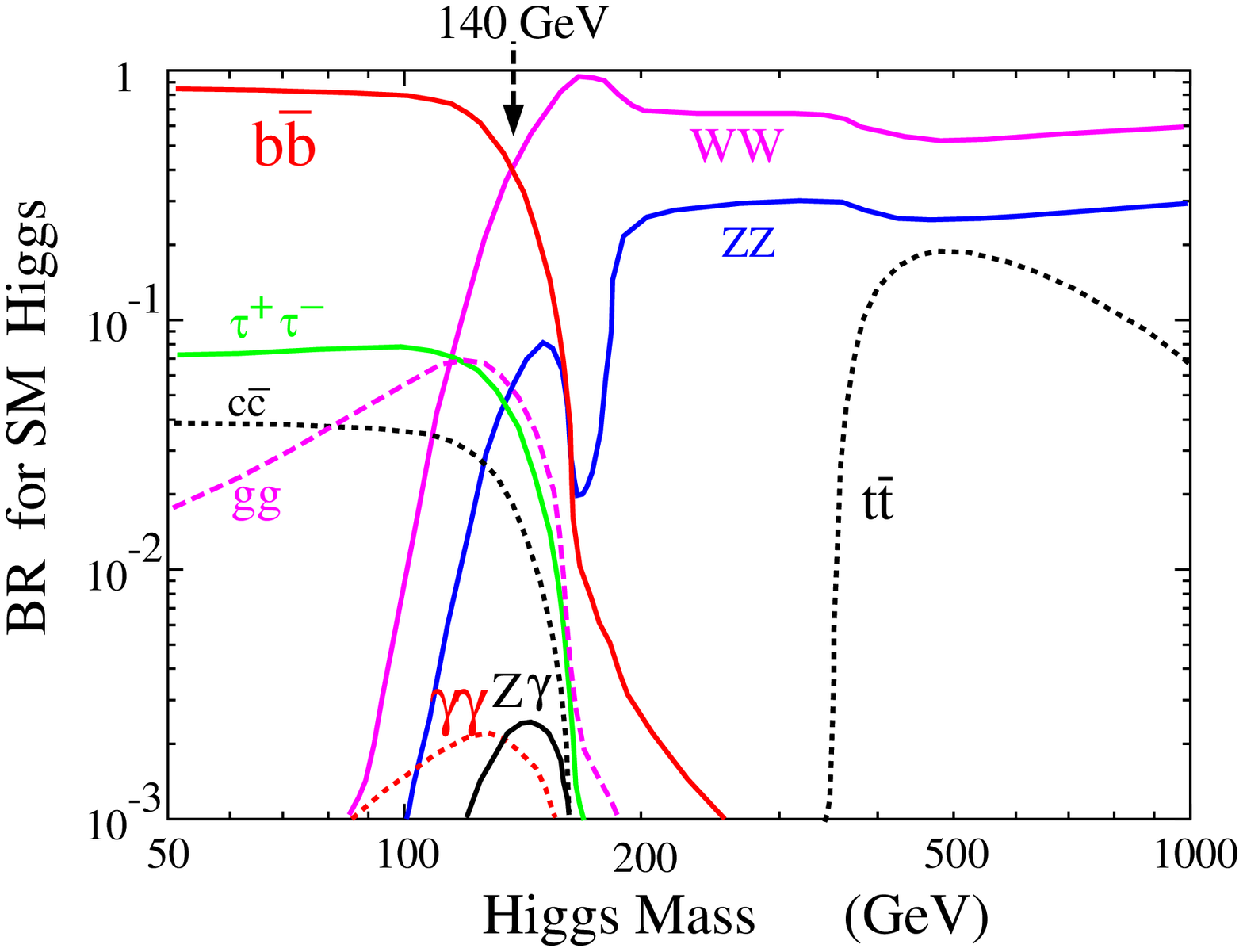}
    \vspace{0.2cm}
  \end{minipage}
  \caption{\label{fig:prod} 
    Left: The main production for a Standard Model Higgs boson at the Tevatron \cite{maltoni}. All three leading production mechanisms, gluon fusion, associated production with a heavy gauge boson and vector boson fusion, plotted in green, red and cyan respectively, are considered in the present analysis. 
    Right: Branching ratios of the Higgs boson as function of the Higgs boson mass.
  }
\end{figure}

\subsection{Signal \& Background Processes}

The main backgrounds of the \hwwll analyses are Drell-Yan (DY), \tt, \wjets and diboson production. While DY can be easily suppressed by requiring large \etmiss~the latter two ones will be the most dominant backgrounds at the final selection stage. In particularly the production of two $W$ bosons results in an identical final state as the main signal process. Mainly the angular correlation of the final state leptons allows for a separation between those processes because the Higgs boson is a spin 0 particle. Instrumental background from \wjets production arises mostly when a jet is mis-identified as lepton. At higher jet-multiplicities where the signal becomes enriched with events produced in VBF, \tt~production gains importance.
The dominant background processes for the hadronic $H \rightarrow WW$ analysis are \wjets and multijet processes. Searches exploiting the $WH$ production mechanism in the same-sign lepton and trilepton final state are as well dominated by instrumental background, e.g charge-mismeasurements or $W+jets/\gamma$ production in which the jet or the photon is mis-identified as lepton.

The simulation of signal and background processes is performed using \textsc{Pythia} and \textsc{Pythia+Alpgen} Monte Carlo (MC) simulations. At \DZero the $WW$ background is reweighted according to \textsc{Sherpa} and CDF uses a \textsc{MC@NLO} simulation. The predictions from MC are normalized to NLO cross section calculations for diboson and DY processes whereas the double top cross section is calculated at NNLL accuracy. The signal processes in associated and gluon-fusion production are normalized to NNLO calculations and NLO for VBF. The normalization of the instrumental \wjets backgrounds is done using data driven methods.

\subsection{Background Discrimination}
All analyses use multivariate discriminator, a commonly used technique in modern High Energy Physics. These techniques are Boosted Decision Trees (BDT), Random Forrest (RF) and Neural Networks (NN). BTDs and RFs are binary tree structures which are relatively insensitive to poorly discriminating variables. NNs are a non-linear statistical data modeling tool which are inspired by the structure of biological neural networks.  These techniques are used to model complex relationships between input variables. The automated learning process is performed using background and signal MC. During this learning process the BDT and RF are adaptively changed by re-arranging the tree structure to optimize signal and background separation. For Neural Networks this optimization is done by 'back-propagation', a algorithm in which the response of individual nodes of the neural networks feed back to alter the internal structure.

\subsection{ \hwwll } \label{sec:hww}

This channel is the classical 'working horse' of high mass Higgs boson searches. The main signal here is specified by an opposite sign lepton pair and large \met. The main signal arises from gluon-fusion production and is expected to show little or no jet activity. About $~5\%$ of the Higgs bosons are produced by VBF and accompanied by jets. 
The \DZero selection requires leptons to be isolated with a transverse momentum of $p_T > 10$\gevc~for muons and $p_T > 15$\gevc~for electrons. Jets are required to have $p_T > 15$\gevc~and $|\eta|<2.4$, their tracks must be associated with the primary event vertex. Details about the object identification and selection can be found in Ref.~\cite{d0_hww}.
CDF selects isolated electrons and muons with at least  $p_T > 18$\gevc. Both collaborations require an offline \met~of at least 20~GeV for the event to pass the selection. Details are to be found in Ref.~\cite{cdf_higgs}.

The various background processes and relative contributions are not independent of lepton flavor, jet multiplicity and the Higgs boson mass. Multijet and instrumental background will strongly differ for different lepton flavors and the \tt~background becomes dominant for \vbf production. By arranging the selected events into bins of orthogonal lepton flavors and jet multiplicities the multivariate methods can be trained and applied to a more well defined phase space resulting in improved signal-to-background ratio. The training is repeated for each analyzed Higgs boson mass in bins of 5\gevcc.

\DZero arranges their events in sub-channels of $e^\pm e^\mp$, $\mu^\pm \mu^\mp$ and $e^\pm \mu^\mp$ events whereas the $e^\pm \mu^\mp$ channels further benefits from splitting into different jet multiplicities. CDF generally split all of their events according to their jet multiplicity (0, 1, or $\ge$2 jets) and  high- or low signal purity, based on the lepton selection used. The sample with two or more jets is not divided into signal purity categories. 
CDF as well uses final states with hadronic taus in the \hww searches. To suppress instrumental background and contributions from the $Z \rightarrow \tau \tau$ peak the selection applies tighter lepton momentum requirements ($p_T>20$\gevc) and applies requirements on the transverse mass of the lepton pair and the angle between lepton pair and \etmiss. The tau identification as CDF is based on 1- and 3-prong tau candidates and employs various requirements on energy/momentum ratios to reduce mis-reconstructions from first or second generation leptons. Details to selection and analysis can be found in Ref.~\cite{cdf_hwwtau}.

For the final selection \DZero uses a BDT approach in the $e^\pm \mu^\mp$ final state and NNs for $e^\pm e^\mp$, $\mu^\pm \mu^\mp$. CDF applies an  NN to all analysis sub-channels but the $\tau$ channel which uses a BDT. For the most sensitive 0-jet bin  Matrix-Elements (ME) (Ref.~\cite{me}) are calculated by CDF. These ME are used as input variable of the NN. Details can be found in Ref.~\cite{d0_hww},~\cite{cdf_higgs}. The final output classifier for the most sensitive sub-channels channels of \DZero and CDF can be found in Fig.~\ref{fig:hww}. The \DZero run period is split into two different epochs, marked by an upgrade of the tracking and triggering system in 2006. The period before the upgrade is called RunIIa and corresponds to 1.1\ifb~whereas the latter period named RunIIb corresponds in this analyses to 5.6\ifb. Figure~\ref{fig:hww} left shows the RunIIb distribution.

\begin{figure}[h]
  \begin{centering}
    \includegraphics[width=0.4\textwidth]{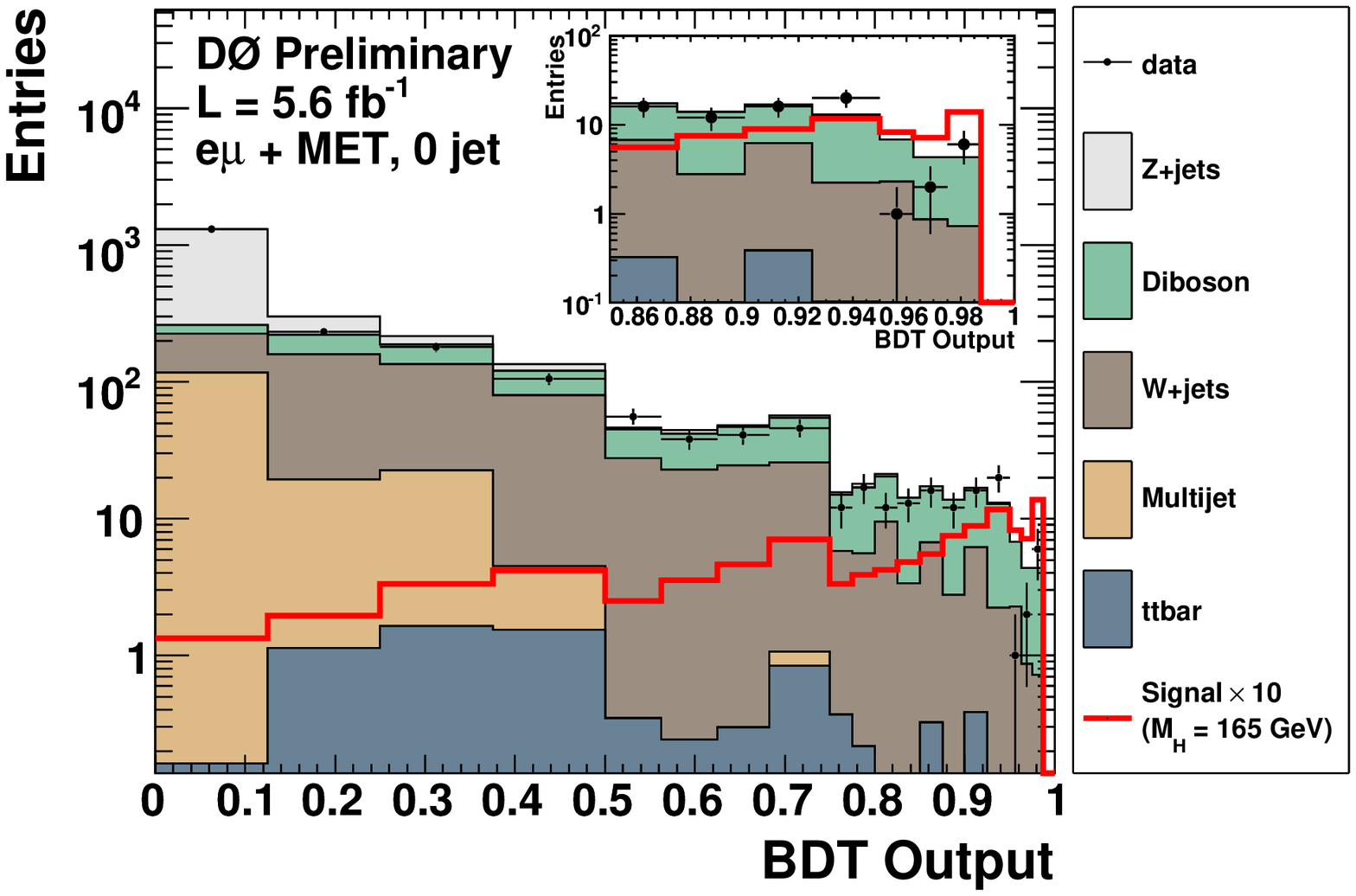}
    \hspace{1.5cm}
    \includegraphics[width=0.4\textwidth]{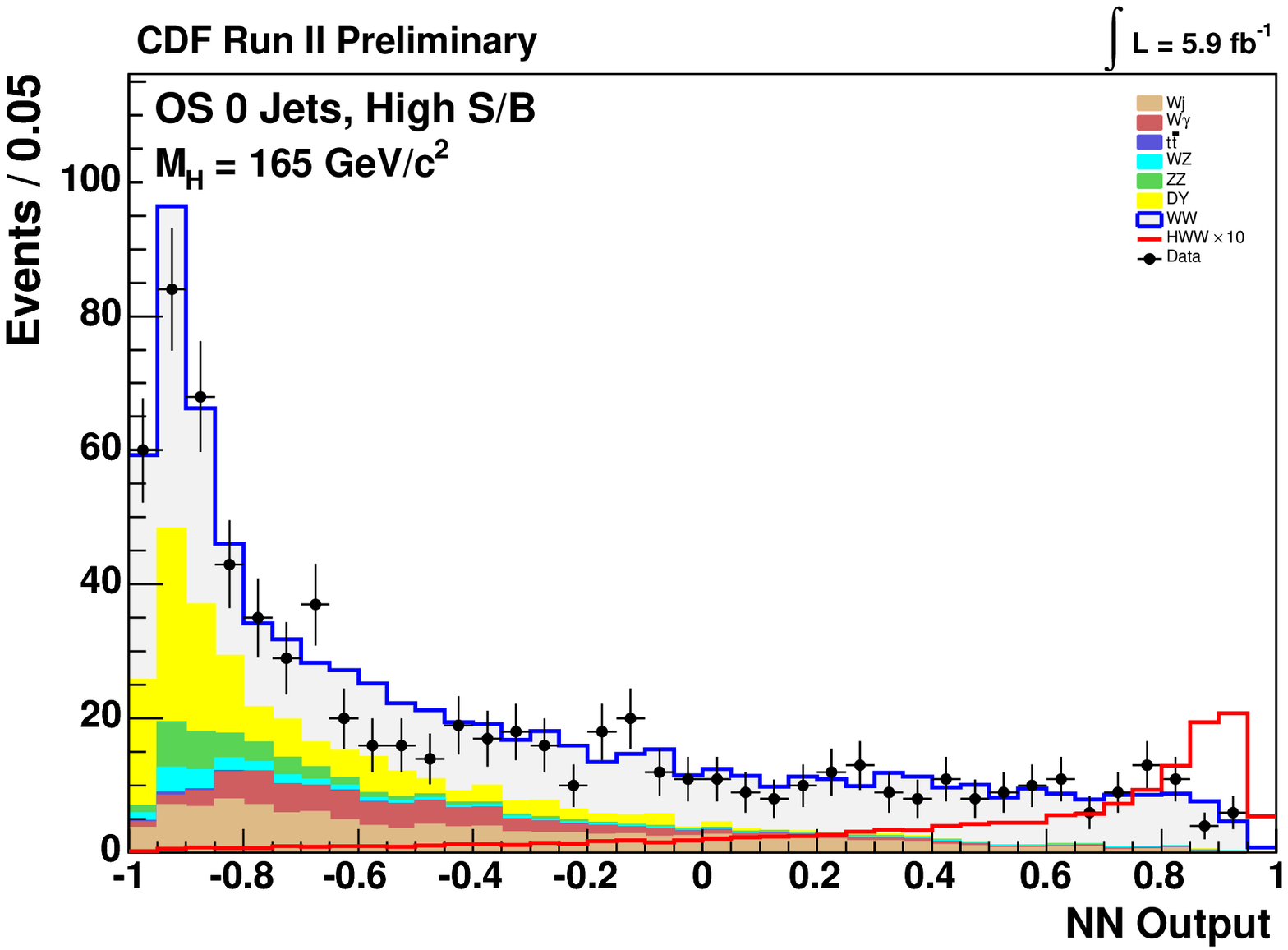}
    \caption
        {\label{fig:hww} \footnotesize{Final output classifier for the most sensitive analysis channels for \DZero (left) and CDF (right). The signal is multiplied by a factor of ten. The \DZero graph corresponds to the run period known as RunIIb. }}
  \end{centering}
\end{figure}

\subsection{  $H \rightarrow W^\pm W^\mp \rightarrow \ell^\pm \nu jj $} \label{sec:hwwhadr}
Unfortunately small branching fractions limit the rates for all-leptonic final states. Decay channels containing a single charged lepton have more background but their rates are also a factor of $\sim 6$ higher than for the all-leptonic states. The analysis presented for the first time this summer considers final state topologies with a single lepton ($e$~or~$\mu$), two or more jets, and \met~arising from \hwwlnujj.

The analysis selects candidates for $W \rightarrow \ell \nu$ decays by requiring $\etmiss > 15$~GeV and the presence of a isolated lepton with $p_T > 15$\gevc. The jets are required to exceed a transverse momentum of $p_T>20$\gevc. To suppress background from multijet events, events are required to have $m_T^W > 40-0.5 \times \met$, where $m_W^T=\sqrt{\left( p_T^\ell p_T^\nu \right)^2 - \left( \mathbf{p}_T^\ell + \mathbf{p}_T^\nu  \right)^2 }$ is the transverse mass of the $W$ boson.
The MJ is background is derived using data driven techniques. In the electron channels two isolation criteria are defined and the MJ background is based on leptons which pass the loose but fail the tight selection. For muons the MJ background is estimated by reversing the isolation criteria. In both cases an event to even weight is measured by examining several topological and kinematic distributions. Details can be found in Ref.~\cite{d0_hadHww}. Figure~\ref{fig:pTW} shows the good agreement achieved despite the challenging background processes.

\begin{figure}[h!]
  \begin{centering}
    \includegraphics[width=0.6\textwidth]{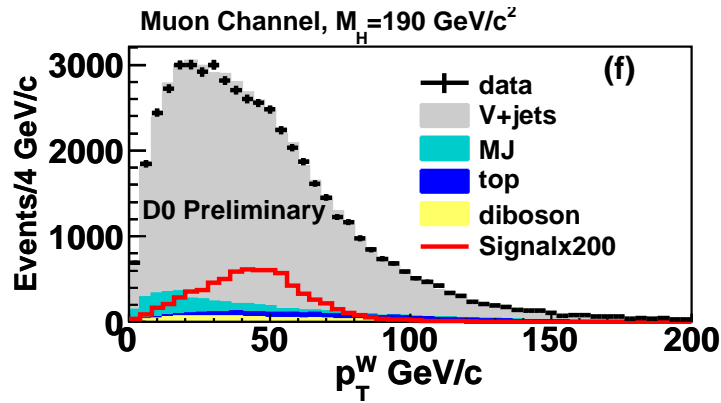}
    \caption
        {\label{fig:pTW} \footnotesize{Transverse momentum of the leptonically decaying $W$ boson in the $\mu$ channel. The background is given by filled histograms, the data by the data points and an signal expection for \mh=165\gevcc~ is indicated by the red histogram. The signal is multiplied by a factor of 200.}}
  \end{centering}
\end{figure}

The final results is based on a random forest method based on a collection of decision trees (DT) which are trained on randomly selected subsamples from both, signal and background MC events. Each DT is trained using $\sim 30$ discriminating variables which were selected using a Kolmogorov-Smirnov test for differences in distributions between signal and background. The final output classifier is shown in Fig.~\ref{fig:hadHww_final}

\begin{figure}[h!]
  \begin{centering}
    \includegraphics[width=0.9\textwidth]{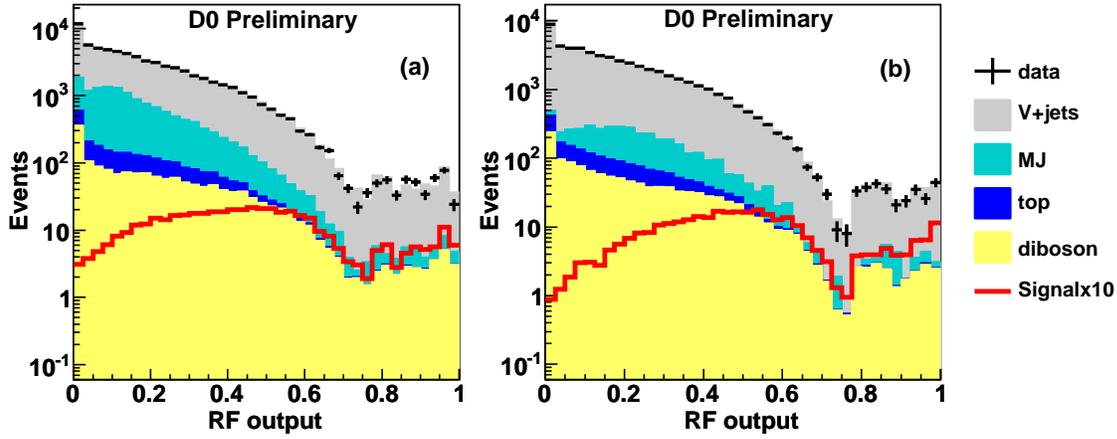}
    \caption
        {\label{fig:hadHww_final} \footnotesize{Random forest outputs for $m_H = 165$\gevcc: for the electron channel (a) and the muon channel (b). The data are
            shown as points with error bars. The background contributions are shown as histograms.}}
  \end{centering}
\end{figure}

For $\mh = 165$\gevcc, the observed and expected limits on the combined cross section for Higgs boson production, multiplied by the branching fraction for \hww are factors of 5.5 and 3.8 larger than the SM value, respectively. These are the first limits on standard-model Higgs boson production examining decays of the Higgs to two bosons having
respective decays into a leptonic and a hadronic final state.

\subsection{ $W/Z+H \rightarrow W/Z+WW \rightarrow \ell^\pm \ell^\mp(\ell) + X$}

Further sensitivity comes from searches in same-sign (SS) dilepton and trilepton final states. These final states utilize mostly the WH/ZH production modes for the high mass Higgs boson searches.

\subsubsection{Same-sign analysis}

Final states with two same sign leptons occur naturally in $VH \rightarrow VWW$ production, when the vector boson ($V= Z,~W$) and one of the W bosons from the Higgs boson
decay leptonically. \DZero requires two leptons with $p_T>15$\gevc, CDF applies the same lepton criteria as described in Sec.~\ref{sec:hww}. To reduce backgrounds in the same-sign final state this base selection has been modified.  In particular events containing forward electrons are not accepted because they have a high charge mismeasurement rate. The central electron candidates are are required to pass tighter isolation requirements. To further reduce the number of photons or jets misidentified as leptons, the $p_T$ requirement for the second lepton is increased from $10$\gevc~to $20$\gevc. Since the decay of the third boson most often results in the production of additional
jets, we also require one or more jets in the final state.
CDF uses a NN and \DZero a DT for the final background and signal discrimination. At $\mh = 165$\gevcc~the \DZero analysis yields an observed (expected) limit of 7.2 (7.0)  on SM Higgs boson production times branching ratio. CDF's analysis results in observed (expected) limit of 6.0 (4.9).

\begin{figure}[h!]
  \begin{centering}
    \includegraphics[width=0.4\textwidth]{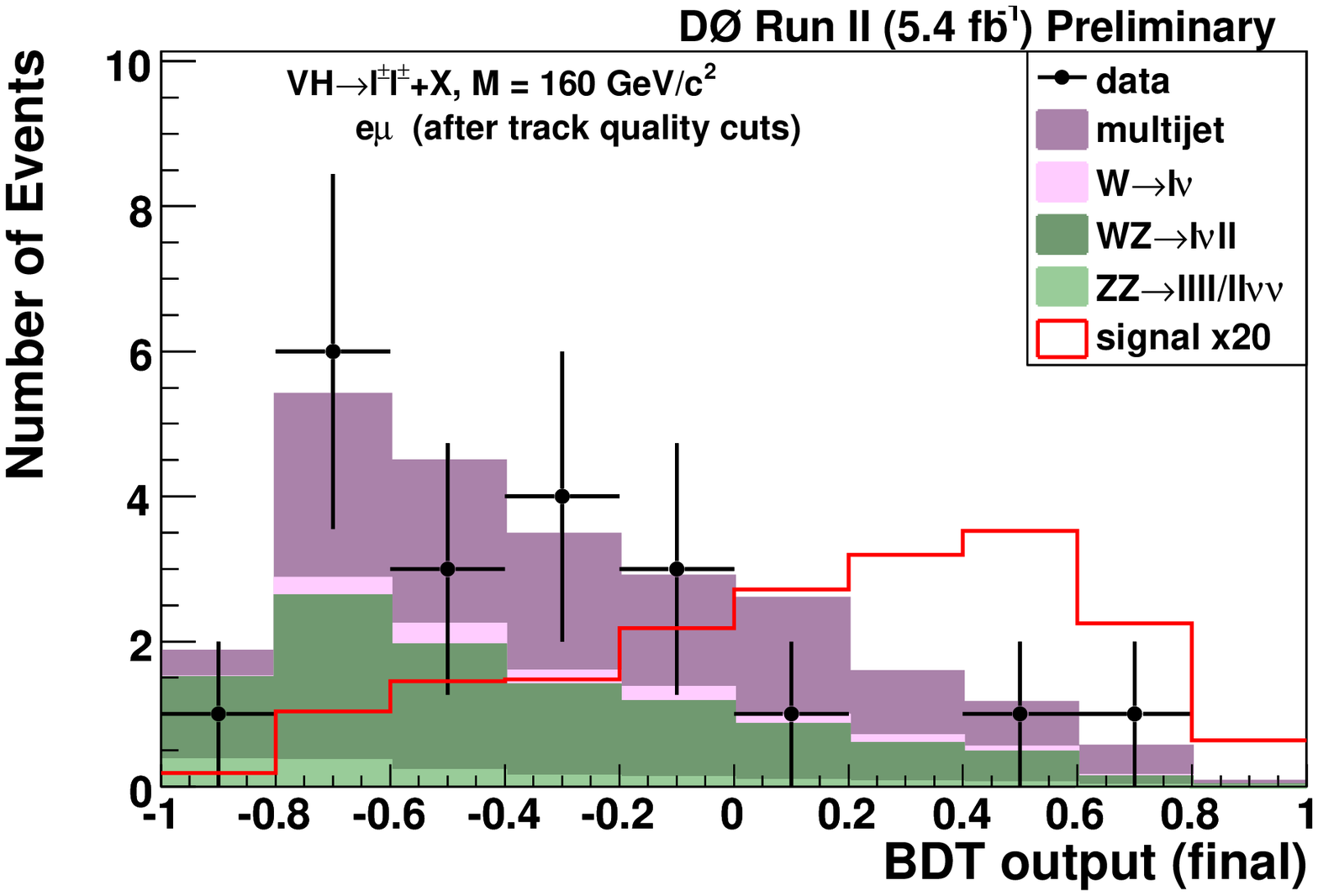}
    \hspace{1.5cm}
    \includegraphics[width=0.4\textwidth]{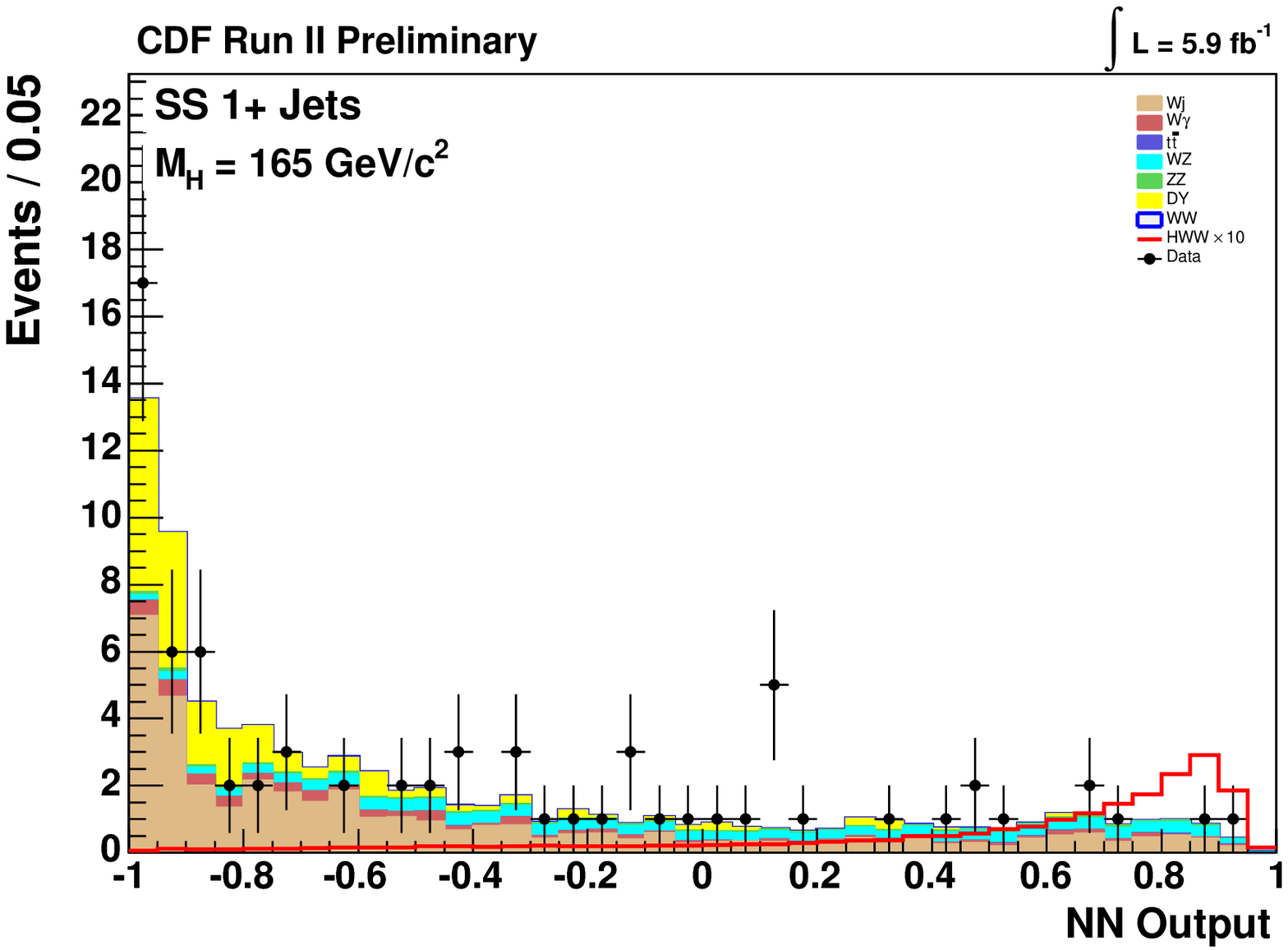}
    \caption
        {\label{fig:hww} \footnotesize{Final output classifier for the most sensitive analysis channels for \DZero (left) and CDF (right). The signal is multiplied by a factor of ten. The \DZero graph corresponds to the run period known as RunIIb. }}
  \end{centering}
\end{figure}

More details can be found in Ref.~\cite{d0_lsgn},~\cite{cdf_higgs}.

\subsubsection{Trilepton analysis}

Another addition performed by the CDF collaboration is the search for potential Higgs boson signal in the trilepton final state as opposed to the dileptonic channels. 
Trilepton events occur naturally in $WH \rightarrow WWW$ production, in the case where all three $W$ bosons decay leptonically, and in $ZH \rightarrow ZWW$ production, 
where the $Z$ boson and one of the $W$ bosons from the Higgs boson decay leptonically while the second $W$ boson decays hadronical. 
The primary background in this search is $WZ$ production, which also can result in a signature of three leptons and
missing energy. At least one lepton is required to satisfy $p_T > 20$\gevc. This requirement is loosened to $p_T > 10$\gevc~for the second and third leptons to increase 
Higgs boson kinematic acceptance. Three channels are considered to better discriminate against the dominant WZ background. These channels depend on the
number of reconstructed jets and whether or not there are two same-flavor ($ee$ or $\mu\mu$) opposite-sign leptons with an
invariant mass that falls within $10$\gevcc~ of the $91$ \gevcc~ $Z$-boson mass. Trilepton events with a same-flavor opposite-sign dilepton pair in the $Z$-mass peak have a Higgs boson signal contribution predominantly from ZH production. Because $ZH$ trileptons  must also contain a high-$p_T$ neutrino $\etmiss > 10$~GeV is required. The likely hadronic decay of the second $W$ boson in $ZH$ trilepton events results in the production of additional jets, so one or more jets are required. 
In case of two or more jets the Higgs boson mass can be reconstructed, thus events with one or two and more reconstructed jets are separated in two analysis channels. Trilepton events without a same-flavor opposite-sign dilepton pair in the $Z$-mass peak have a Higgs boson signal contribution predominantly from $WH$ production. Because most $WH$ trilepton events contain three neutrinos $\etmiss>20$~GeV is required. Since $WH$ trilepton events contain no jets at leading order no requirement on the jet multiplicity is applied.
Figure~\ref{fig:trilep} shows NN templates used for the final measurement in this channel. 

\begin{figure}[h!]
  \begin{centering}
    \includegraphics[width=0.4\textwidth]{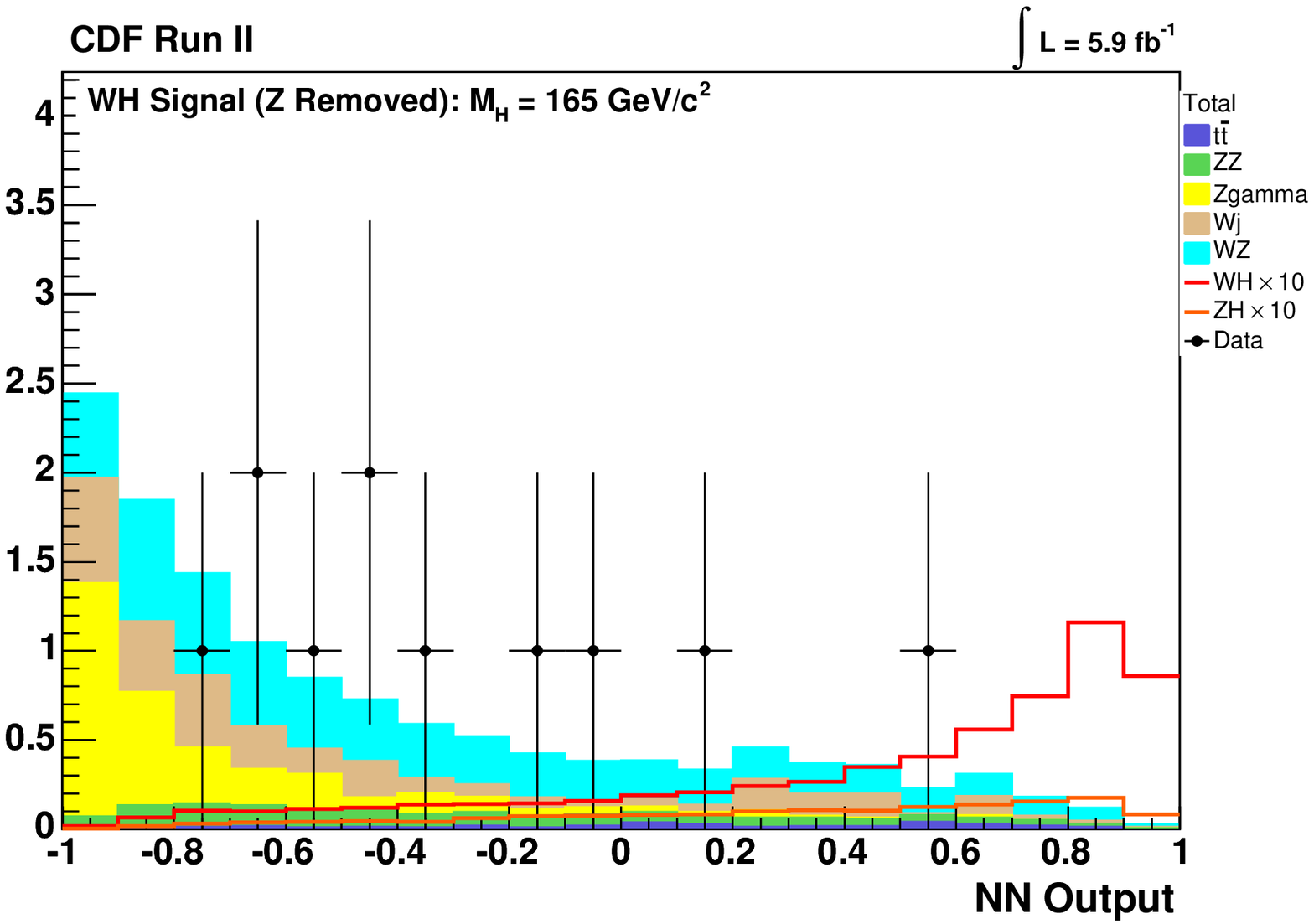}
    \hspace{1.5cm}
    \includegraphics[width=0.4\textwidth]{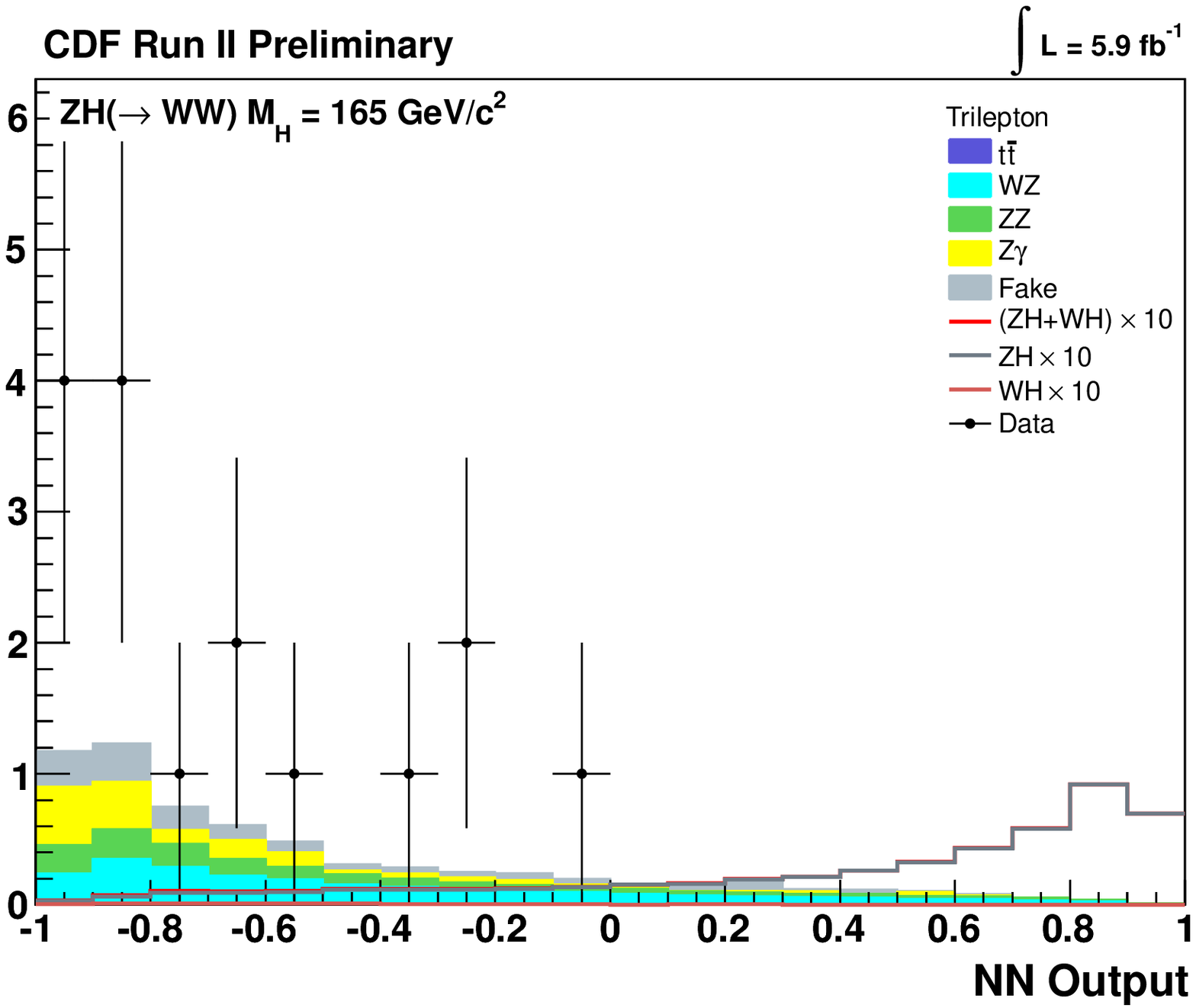}
    \caption
        {\label{fig:trilep} \footnotesize{Final output classifier for the $WH$ search channel 'trilepton outside Z' (left) and the $\ge 2$jet 'trilepton in Z' channel (right).}}
  \end{centering}
\end{figure}

The observed (expected) limit of the 'Trilepton outside Z' channel at \mh=165\gevcc~is 7.9 (7.4). For the same Higgs boson mass the 'Trilepton in Z' 1 jet channel yields to 36.4 (31.8) and the 
$\ge 2$ jet channel  10.4 (9.2).

\subsection{Systematic Uncertainties} 

Systematic uncertainties differ between experiments and analyses, and they affect the rates and shapes of the predicted
signal and background in correlated ways.  The combined results incorporate the sensitivity of predictions to  values of nuisance parameters,
and include correlations, between rates and shapes, between signals and backgrounds, and between channels within one collaboration and between both collaborations. We correlate various systematics like the uncertainties on the integrated luminosity and production cross sections. Other systematics like for example the $b$-tagging uncertainty or multijet background cannot be correlated due to the different techniques employed, but these systematics are correlated within each experiment. All systematic uncertainties arising from the same source are taken to be correlated between the different backgrounds and between signal and background. The tyical uncertainties for signal and background processes range from $\sim15-20\%$.

\section{RESULTS}

Using the combination procedures outlined in Ref.~\cite{pflh}, limits on SM Higgs boson production $\sigma \times B(H\rightarrow X)$ in
\pp~collisions at $\sqrt{s}=1.96$~TeV are extracted.  To facilitate comparisons with the standard model and to accommodate analyses with
different degrees of sensitivity the results are presented in terms of the ratio of obtained limits to cross section in the SM as a function of the 
Higgs boson mass. A value of the combined limit ratio less than or equal to one indicates that that particular Higgs boson mass is excluded at the 95\% C.L. 

The combinations of channel of the individual experiments, yield to limits of 1.1~(1.0) for CDF and 1.0~(1.1) for \DZero at $m_{H}=165$\gevcc~as ratios of 95\% C.L. observed (expected) limits to the SM  cross section.
For the combined \DZero and CDF analysis we obtain the observed (expected) values of 0.69 (0.73) at $m_{H}=165$\gevcc. We exclude at the 95\% C.L. the production of a standard model Higgs boson with mass between 158 and 175\gevcc. The expected exclusion ranges from 156 to 173\gevcc.
The ratios of the 95\% C.L. expected and observed limit to the SM cross section are shown in Fig.~\ref{fig:comboRatio} for the combined CDF and D\O\ analyses.

\begin{figure}[h!]
  \begin{centering}
    \includegraphics[width=15.5cm]{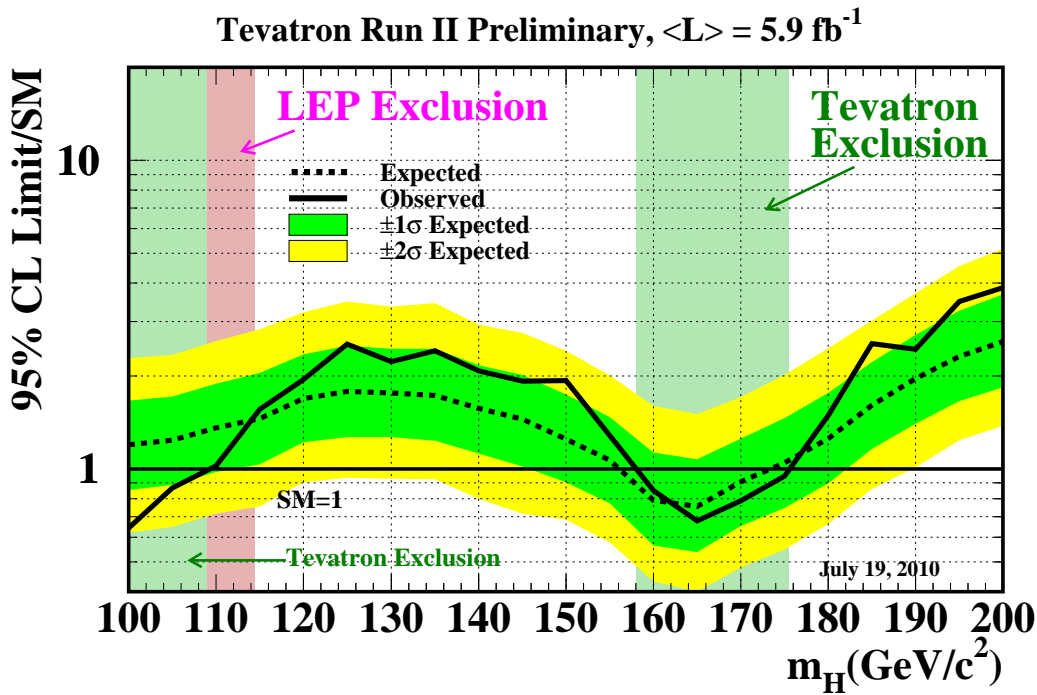}
    \caption{
      \label{fig:comboRatio}
      \footnotesize{Observed and expected (median, for the background-only hypothesis)
        95\% C.L. upper limits on the ratios to the SM cross section, as functions of the Higgs boson mass for the combined CDF and D\O\ analyses.
        The limits are expressed as a multiple of the SM prediction for test masses (every 5\gevcc) for which both experiments have performed dedicated
        searches in different channels. The points are joined by straight lines for better readability.   The bands indicate the
        68\% and 95\% probability regions where the limits can fluctuate, in the absence of signal. The limits displayed in this figure
        are obtained with the Bayesian calculation.}
    }
  \end{centering}
\end{figure}

In summary, analyses and the combination of all available CDF and D\O\ results on high mass SM Higgs boson search, based on luminosities ranging from 5.4 to 6.7 fb$^{-1}$ have been presented. Compared to previous combination, new channels have been added and previously used channels have been reanalyzed to gain sensitivity. The 95\% C.L. upper limits on Higgs boson production are a factor of 0.69 times the SM cross section for a Higgs boson mass of $m_{H}=$(165)\gevcc.  These results extend significantly the individual limits of each collaboration and the previous combination. The mass range excluded at 95\% C.L. for a SM Higgs has been extended to  $158<m_{H}<175$\gevcc. The sensitivity of the combined search is expected to grow substantially in the near future with the additional luminosity already recorded at the Tevatron, additional improvements of analysis techniques and consequent use of more analysis channels as presented here.

\begin{acknowledgments}
The authors wishes to thank all analyzers of the presented analyses and the \DZero and CDF Higgs conveners for their support in preparing the presentation and these proceedings.
\end{acknowledgments}

\bibliography{proceedings}{}
\bibliographystyle{unsrt}

\end{document}